# Time transfer through optical fibers over a distance of 73 km with an uncertainty below 100 ps

M Rost[1,*], D Piester[1,**], W Yang[2], T Feldmann[1,***], T Wübbena[3] and A Bauch[1]

[1]Physikalisch-Technische Bundesanstalt (PTB), Bundesallee 100,

38116 Braunschweig, Germany

[2]National University of Defense Technology (NUDT), Changsha, PR China

[3]Institut für Quantenoptik (IQ), Leibniz Universität Hannover, Germany

* Present address: DLR, Institut für Verkehrssystemtechnik, Braunschweig, Germany

** Contact: dirk.piester@ptb.de

*** Present address: TimeTech GmbH, Stuttgart, Germany

**Abstract**

We demonstrate the capability of accurate time transfer using optical fibers over long distances utilizing a dark fiber and hardware which is usually employed in two-way satellite time and frequency transfer (TWSTFT). Our time transfer through optical fiber (TTTOF) system is a variant of the standard TWSTFT by employing an optical fiber in the transmission path instead of free-space transmission of signals between two ground stations through geostationary satellites. As we use a dark fiber there are practically no limitations to the bandwidth of the transmitted signals so that we can use the highest chip-rate of the binary phase-shift modulation available from the commercial equipment. This leads to an enhanced precision compared to satellite time transfer where the occupied bandwidth is limited for cost reasons. The TTTOF system has been characterized and calibrated in a common clock experiment at PTB, and the combined calibration uncertainty is estimated as 74 ps. In a second step the remote part of the system was operated at Leibniz Universität Hannover, Institut für Quantenoptik (IQ) separated by 73 km from PTB in Braunschweig. In parallel, a GPS time transfer link between Braunschweig and Hannover was established, and both links connected a passive hydrogen maser at IQ with the reference time scale UTC(PTB) maintained in PTB. The results obtained with both links agree within the 1-σ uncertainty of the GPS link results, which is estimated as 0.72 ns. The fiber link exhibits a nearly 10-fold improved stability compared to the GPS link, and assessment of its performance has been limited by the properties of the passive maser.

## 1. Introduction

Remote comparisons are indispensible in the metrology of time and frequency, and the recent progress in the accuracy of optical frequency standards [1] motivated several studies on optical frequency transfer through optical fibers [2, 3, 4]. This has to be distinguished from other objective targets such as transfer of standard frequencies modulated on optical carriers [5, 6] and accurate clock (time) comparisons using optical fibers [7, 8, 9]. In the latter cases often the term "time transfer" is used although a calibration of propagation delays is left undone so that the results demonstrate the precision and in some cases the reproducibility of the employed systems, but not necessarily the accuracy to compare epochs of the two (or more) clocks involved.

In previous experiments [10] we demonstrated time transfer using optical fibers (TTTOF) on the PTB campus with a reproducibility of better than 100 ps. The two-way technique employed guarantees cancellation of all long-term fiber variations due to the symmetry of the signal propagation path, as it was demonstrated before for frequency transfer [11]. Subsequently we will provide a full calibration budget for this type of time transfer. The previously developed system will be used to synchronize a remotely generated time-scale at a distant "satellite" time laboratory on PTB's campus [12], serving two-way satellite time and frequency transfer (TWSTFT) terminals and in the future one ACES microwave link ground terminal (ACES stands for Atomic Clock Ensemble in Space, [13]).

In this work we report on the operation of the TTTOF system in parallel to a calibrated GPS time-link over a distance of 73 km between the Physikalisch-Technische Bundesanstalt (PTB) and the Institut für Quantenoptik (IQ) at Leibniz Universität Hannover. The dedicated dark telecommunication fiber (SMF-28) installed between the two sites was used for experiments on optical frequency transfer before [14]. Details of the fiber route can be found in [15]. Section 2 of this paper contains an overview of the basic concept of two-way time transfer and a description of the experimental setup. In section 3 we detail the calibration procedure of the TTTOF equipment before we present in section 4 the results obtained. A conclusion and an outlook on further work are given in section 5.

## 2. Time transfer using the two-way method

*2.1 Experimental set-up*

TWSTFT has been used for many years for time transfer via geostationary satellites between various National Metrology Institutes (NMIs) participating in the generation of International Atomic Time (TAI) by the Bureau International des Poids et Mesures (BIPM) [16]. It is based on the exchange of pseudorandom noise (PRN) binary phase-shift keying (BPSK) modulated carrier signals through geostationary telecommunication satellites. The signals are travelling along reciprocal paths, and thus to first order propagation delays due to ionosphere, troposphere and to changing satellite position are suppressed. The phase modulation is synchronized with the local clock's 1 PPS (one pulse per second) output. Each station uses a dedicated PRN for its BPSK sequence in the transmitted signal. The receiving equipment is able to generate the BPSK sequence of the remote stations and to reconstitute a 1 PPS tick from the received signal. This is measured by a time-interval counter (TIC) with respect to the local clock. Instead of translating the PRN modulated signal at 70 MHz to the Ku-band for transmission, it is in the present case modulated onto an optical carrier which is then transmitted through a fiber.

In figure 1 a schematic view of the experimental setup is shown. Both parts of the transfer system include a Satellite Time and Ranging Equipment (SATRE) modem manufactured by TimeTech GmbH, Stuttgart, Germany. The modems are supplied with 10 MHz and 1 PPS reference signals representing the time scales or clocks at each side. In the remote setup a frequency distribution amplifier (FDA) and a pulse distribution amplifier (PDA) provide reference signals for the accompanying GPS timing receiver. The chip rate of the modem output signals can be configured in several steps from 1 to 20 Mcps. For best performance we use the maximum chip rate of 20 Mcps in our experiment, resulting in the lowest instability of the comparison data. The output signals are modulated onto optical signals at 1550 nm by using electro-optical transmitters (E/O) from Linear Photonics (IFL TimeLink). Optical circulators are used to combine transmitted and received signals at both ends of the fiber. Optical isolators at both sides protect the E/O against back-scattered light from the circulators. The two optical signals are transported through one single-mode fiber in opposite directions whereby the signals in both directions should be influenced identically during their passage so that the two-way data analysis, described in the following section removes such effects to first order. At each side, an opto-electrical receiver (O/E) converts the signal again to 70 MHz electrical signal, which is subsequently fed to the receiving part of the modems.

In this experiment we wanted to avoid any additional optical amplifier and thus its contributions to the transfer uncertainty [7, 17], so that the maximum distance that could be bridged was about 100 km for the following reason. The E/O converter accepts a maximum electrical input power of 0 dBm, and the input to the modem at the end has to be larger than –55 dBm. The optical output power of the E/Os (5 dBm maximum) is reduced by about 0.2 dB/km inside the fiber and by additional losses in connectors, in total by 27 dB. Previous experiments showed that the internal modem delay is power dependent, so the electrical input signal power to the modems has to be adjusted to a constant value via a variable attenuator in the optical path, independent of the length of the fiber used.

## 2.2   Definition of the observables

The following formulas describe the two-way time transfer through the fiber link as illustrated in figure 1. Here we use indices L and R for quantities referring to the local and remote set-up, respectively, which reflects the situation of time transfer from the laboratory to a remote location. The modems perform several functions: 1 PPS signals synchronous with the transmitted PRN signal and with the received PRN signal are generated, designated 1PPSTX and 1PPSRX, respectively. Two measurements are provided: 1PPSTX – 1PPSRX, designated as *TD*, and 1PPSREF – 1PPSTX, designated as *REFDELAY*. The following two relations are valid (using designations from figure 1):

$$TD_L = TA_L - CAB_L - REFDELAY_L - TA_R + CAB_R + REFDELAY_R + PROP_{R,L} \qquad (1)$$

$$TD_R = TA_R - CAB_R - REFDELAY_R - TA_L + CAB_L + REFDELAY_L + PROP_{L,R}. \qquad (2)$$

*CAB* represents the cable delay between the local time scale reference point and the input to the modem. *PROP*$_{L,R}$, e. g., is the sum of the signal delay in the local setup, *TX*$_L$, of *SP*$_{L,R}$, and of the signal delay in the receive branch of the remote setup, *RX*$_R$. *SP*$_{L,R}$ is the transmission path delay from the local setup to the remote setup, and *SP*$_{R,L}$ is the delay of the signal using the same path in the opposite direction. For a single optical fiber connection between the two sites *SP*$_{L,R}$ = *SP*$_{R,L}$ is valid in good approximation. The two measurements performed finally provide the quantity of interest, $TA_L - TA_R$, where $TA_L$ and $TA_R$ represent the local and the remote clock signal (1 PPS), respectively:

$$TA_L - TA_R = \tfrac{1}{2}(TD_L - TD_R) + \{CAB_L\} - CAB_R + \{REFDELAY_L\} - REFDELAY_R +$$

$$\{\tfrac{1}{2}(PROP_{L,R} - PROP_{R,L})\}. \tag{3}$$

The quantities inside curled brackets in (3) are considered as constants. The first two are constant as long as the local set-up remains unchanged. The third one depends on the constant internal delays in the equipment and the condition $SP_{L,R} = SP_{R,L}$. The determination of these terms represents the "calibration" of the time transfer set-up. To this end, the two setups are operated side-by-side in a so-called common clock configuration resulting in $TA_L = TA_R$, thus:

$$0 = \tfrac{1}{2}(TD_L - TD_R) - CAB_R - REFDELAY_R + \{CAB_L + REFDELAY_L + \tfrac{1}{2}(PROP_{L,R} -$$

$$PROP_{R,L})\}. \tag{4}$$

Based on (4), we introduce the quantity *CALR* through

$$CALR = [-\tfrac{1}{2}(TD_L - TD_R) + CAB_R + REFDELAY_R]_{CC} = \{CAB_L + REFDELAY_L + (PROP_{L,R} -$$

$$PROP_{R,L})\}, \tag{5}$$

where the index CC indicates that all quantities inside the squared brackets are determined during the common-clock set-up. Having determined this value, the set-up can be installed at its remote destination, and can provide the difference between two different time scales, $TA_L - TA_R$ as

$$TA_L - TA_R = [\tfrac{1}{2}(TD_L - TD_R) - CAB_R - REFDELAY_R]_{RR} + CALR, \tag{6}$$

without the knowledge of the individual signal delay contributions from (4). Quantities inside $[…]_{RR}$ are determined at the remote site, and numerical values are usually different from those in $[…]_{CC}$ as cables of different length are presumably connecting the local reference signals to the remote setup on both sites. The *REFDELAY* values are (non-zero) constants during each measurement campaign as the input signals (1 PPS and 10 MHz) to both modems originate from the same sources.

## 3. Characterization of statistical and systematic measurement uncertainty

For the initial calibration, both setups were operated for a couple of days in a common clock mode, as described before, connected to UTC(PTB) as a common time and frequency reference. In figure 2a the configuration is shown. For all delay measurements one TIC with a constant trigger level and a 50 Ω input impedance was used. The delay $CAB_R$ between the UTC(PTB) reference point and the input to the remote setup was measured by subsequently connecting UTC(PTB) and 1PPSREF to one input of the TIC while at the second input an auxiliary 1 PPS (phase coherent to UTC(PTB)) was permanently connected. We found an uncertainty of this reference connection of only $U_{REF} = 38$ ps (consisting of statistical and systematic contributions $u_A = 28$ ps and $u_B = 25$ ps, respectively). The additional systematic term for time interval measurements specified by the manufacturer for the time interval counter ($u_B = 0.5$ ns) need not be taken into account because it cancels out. The uncertainty of the *CALR* value is then the quadratic sum of $U_{REF}$ and statistical contributions from the measurements ½ ($TD_L - TD_R$) and $REFDELAY_R$, respectively. During this calibration phase the TTTOF system was operated with a short fiber connecting the local and the remote setup.

In previous experiments the reproducibility of the set-up against power on-off cycles as well as the dependence of measurement results on the length of the fiber used in common-clock mode had been investigated, as such dependence would need to be accounted for in the uncertainty budget [12]. Figure 3 shows results of raw clock differences[1] $RCD = (TD_L - TD_R)/2$ of sequential operation with a short fiber connection and a longer fiber connection with fibers of various lengths on a cable drum. If the assumption $SP_{L,R} = SP_{R,L}$ is valid we would expect no significant variation of the raw clock differences when we change the length of the fiber. However, reflections along the (optical) signal path may cause multipath effects having a possible impact on the overall accuracy. We found a standard deviation (*SD*) of 17 ps. Even better reproducibility had been shown in previous power on-off cycles (*SD* = 6.2 ps) and sequential operation with a 2-m fiber and a 2-km test loop (*SD* = 6.0 ps).

At the local site in Braunschweig the setup was operated simultaneously with a travelling GPS receiver (TR) setup, which had been used to calibrate other time links before [18, 19]. The TR consists of a receiver type GTR50 manufactured by DICOM together with a Stanford Research Systems SR620 TIC. The reference receiver at PTB is an Ashtec Z12-T (acronym PTBB). For the evaluation of the carrier phase, the NRCan-PPP software was used[2] [20]. Shipping the TR after all calibration parameters had been determined, together with the TTTOF setup, allowed us to compare the performance of both time transfer systems as shown in figure 2b. At IQ both devices were connected to a passive hydrogen maser PHM (PTB's property), which had been operated there for quite some time. The electrical output power of the modems and the attenuator in the optical fiber-way were adjusted in such a way, that both modems received the signals of each others with the same signal power as in the initial common clock installation.

The analysis of the stability was performed before and after the TTTOF experiment, so that any changes to the system during the transportation could be detected. Figure 4a shows a phase-plot over a five-day period after the return to PTB. The peak-to-peak variations of raw clock differences lie in a narrow band of 60 ps, except for an artifact at Modified Julian Date MJD 55862.3. The reduced noise between MJD 55863 and 55865 may result from quiet weekend-days in the laboratory. In figure 4b the TDEV for the five-day data set and a one-day measurement period at MJD 55863, respectively, are displayed. The lowest instability can be found for τ between 100 and 1000 seconds. Therefore, and because the used GPS-PPP-data (for details see [18] and references therein) are available every 300 s we chose this as the averaging time of all further analysis.

---

[1] As $TD_L$ and $TD_R$ are based on 1PPS measurements, they are provided by both modems every second and recorded by two computers connected to both modems, respectively.
[2] RINEX data were processed using the NRCan CSRS-PPP software package along with rapid IGS products as precise satellite clock and ephemeris input. The software calculates the position and clock offset for a receiver based on a Kalman-filter algorithm. The parameters for the troposphere model and filter boundary conditions were adjusted in the same way as they were found to be optimal for time transfer by the BIPM.

The TDEV values in figure 4 are always below 6 ps, and as low as 800 fs at averaging intervals of 300 s. Nevertheless in our uncertainty budget we assigned a 6 ps statistical contribution for the raw clock difference measurements. Another uncertainty contribution is due to the power dependence of the measurement results. We showed that it is below 3.4 ps if the power is kept constant to +/– 1 dB, and we were able to keep the power level adjusted to < 0.1 dB.

Due to schedule constraints only a one day common clock measurement was possible before the shipping of the instruments to Hannover, and the mean over this period exhibits a difference from the data shown in figure 4 which amounts to 40 ps. This is larger than the typical reproducibility and stability values obtained in the laboratory as discussed in this section. As there is no explanation right at hand, we include this value of 40 ps in the uncertainty budget, but we are convinced that a repetition of the exercise will either give evidence of an unknown weakness in the set-up or provide a much lower difference. It is evident that this uncertainty contribution needs to be addressed in future time transfer exercises.

The *REFDELAY* values are measured by built-in TICs of the TWSTFT modems. The corresponding statistical uncertainty is mainly limited by the measurement noise introduced by the TICs. Values vary from 10 ps to 40 ps, depending on the device as well as on the 1 PPS shape.

**4. Results of time transfer bridging 73 km**

The TTTOF link was operated for several days in October 2011 between Braunschweig and Hannover. The following dataset was determined from MJD 55848 to 55851. Figure 5 shows the results of the measurement after all calibrations and corrections were attached to the data. The upper graph shows the plain measurement values after adding all calibration constants. The values of the TTTOF (red line) show a much smoother trend than the GPS measurements (blue line).

The graph in figure 5b represents the double difference of TTTOF and GPS data. One can distinguish different signatures of noise throughout the few days. Because of the superior stability of TTTOF they have to be attributed to the conditions of GPS signal reception at both sites. GPS signal recording failed during 15 minutes at the beginning of MJD 55850. The PPP algorithm is sensitive to such data loss as it causes a restart of the calculation. Some parameters are re-calculated leading to the oscillation which can be seen in figure 5b starting with MJD 55850.0. Data taken thereafter (grey-shaded) were not used for determining accuracy and stability. In figure 5c we overlay the residuals to least-squares linear fits individually made to the two data sets. The residual quadratic function can be well explained with the frequency drift of the passive maser.

In figure 6, time instability in terms of time deviation (TDEV) for local and remote links is shown. Basically one notes that the GPS PPP time transfer noise is dominant, and that the 73 km distance between the two sites causes an increase in TDEV by about a factor of 2 for averaging times longer than 1000 s. One also notes that TTTOF over 73 km allows the assessment of the maser instability with almost no noise contribution.

Table 1 contains the uncertainty budget for TTTOF calculated from the individual contributions mentioned in the sections before. The combined uncertainty is a factor 10 lower than standard values stated for GPS and TWSTFT [16, 19]. The mean value calculated from the data in figure 5b is 0.51 ns which is well within the 1-σ combined uncertainty of the calibration delay of the GPS travelling receiver of 0.72 ns. The latter consists of a statistical and systematic contributions of $u_A = 0.24$ ns and $u_B = 0.68$ ns, respectively. The main components of the latter are the closure (0.31 ns) and estimated contributions from multipath, troposphere and phase ambiguities (each 0.3 ns), see [18, 19] for details.

## 5. Summary and conclusion

We temporarily established a time transfer link using a dark telecommunication fiber bridging the distance between PTB in Braunschweig and IQ, Leibniz Universität Hannover. Accuracy and stability of the time link were analyzed and an uncertainty budget was calculated. Our technique is capable of improving time transfer and clock comparison on intermediate distances up to 100 km compared to satellite-based comparisons. We have shown that a true time transfer accuracy better than 100 ps can be achieved. In the future we plan to establish more regular comparisons of this kind which could be termed "long-baseline common-clock GPS comparisons" and study the apparent noise, to be attributed to the GPS link, as a function of satellite geometry, troposphere conditions and receiver type.

As a next step we will study locally the effect of a bi-directional optical amplifier on the transfer stability and then establish a closed loop experiment over 146 km with the two set-ups connected to a common clock. This will allow a better assessment of the ultimate performance of this type of time transfer.

**Acknowledgments**

The authors thank Harald Schnatz and Gesine Grosche of PTB for helpful discussions and support when establishing the fiber connection to Hannover and acknowledge Wolfgang Schäfer, TimeTech GmbH, for providing one SATRE modem on loan base. The authors also thank Stefan Weyers for providing fountain clock data and Miho Fujieda, NICT, for her valuable input. Wenke Yang thanks China Scholarship Council for funding her one year scholarship at PTB.

**Disclaimer**

The Physikalisch-Technische Bundesanstalt as a matter of policy does not endorse any commercial product. The mentioning of brands and individual models seems justified here, because all information provided is based on publicly available material or data taken at PTB and it will help the reader to make comparisons with own observations.

Table 1: Uncertainty Budget: The values "$u_A$ common clock difference" were determined during the calibration of the system; the values "$u_A$ link" were measured during the experiment.

| | uncertainty contribution | uncertainty value (ps) | |
|---|---|---|---|
| $u_A$ common clock difference | RCD | 6 | |
| | $REFDELAY_L$ | 36 | |
| | $REFDELAY_R$ | 10 | total $u_A$ |
| $u_A$ link | RCD | 6 | 49 |
| | $REFDELAY_L$ | 29 | |
| | $REFDELAY_R$ | 10 | |
| $u_B$ | $CAB_R$ calibration | 38 | total $u_B$ |
| | closure | 40 | 56 |
| | power variation | 3 | |
| total uncertainty | | 74 | |

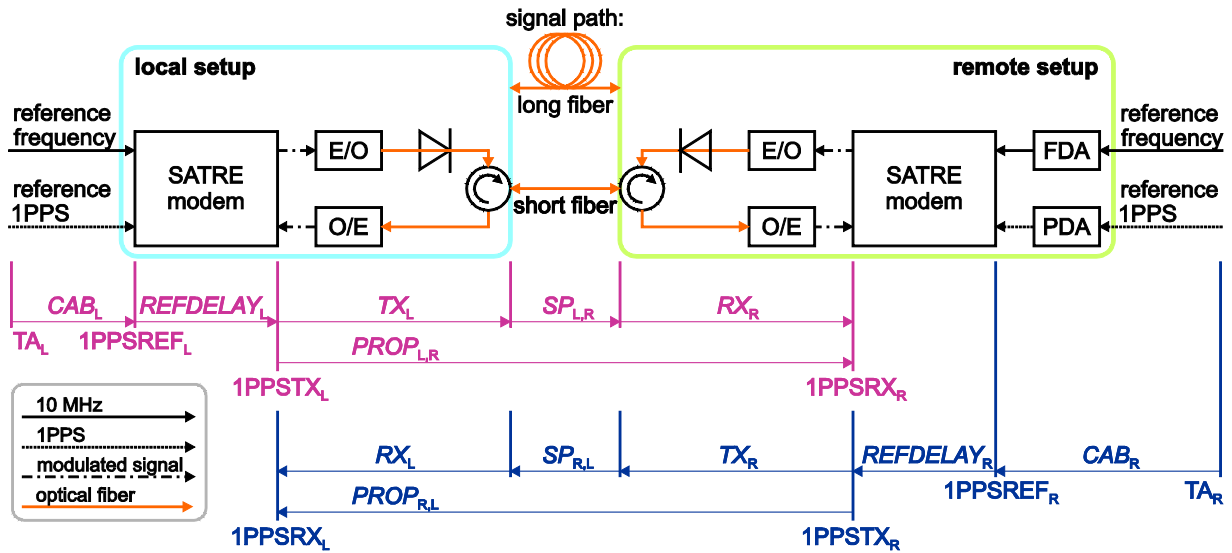

Figure 1: Schematic view of the configuration of our TTTOF experiment. The designations of equipment and measurement quantities are detailed in section 2.1 and 2.2, respectively.

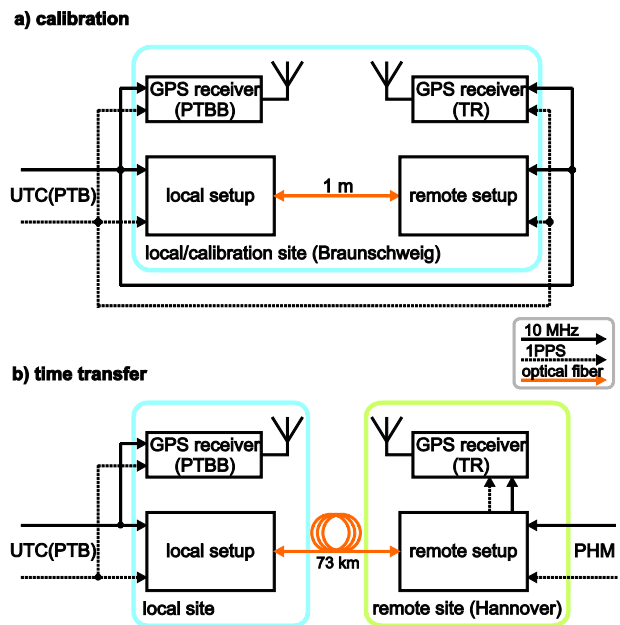

Figure 2: a) Schematic view of calibration of TTTOF system and parallel used GPS calibration system, b) time transfer between Braunschweig and Hannover using both systems in parallel.

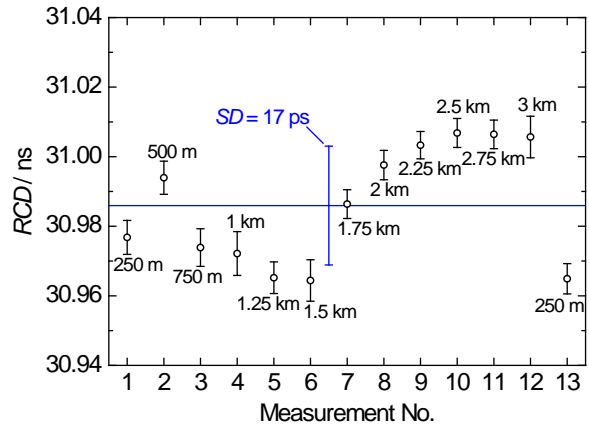

Figure 3: Sequence of raw clock differences $RCD = (TD_L - TD_R)/2$ measurements using different fiber length between 250 m and 3 km, fibers on a cable drum.

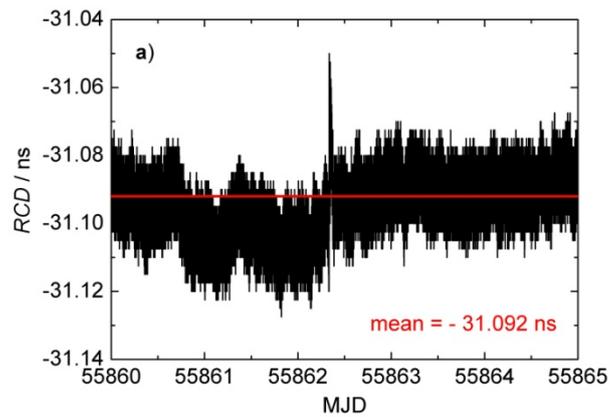

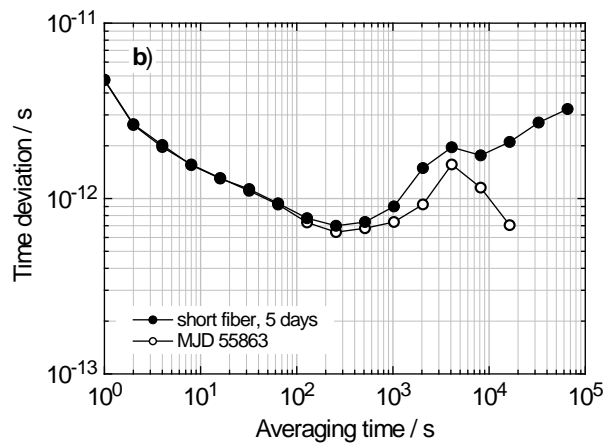

Figure 4: Stability of Common Clock Difference: a) phase of raw clock differences $RCD = (TD_L - TD_R)/2$, b) instability of the 5-day dataset (full dots) and of the data taken during MJD 55863 (open dots). MJD 55860 corresponds to 26 October 2011.

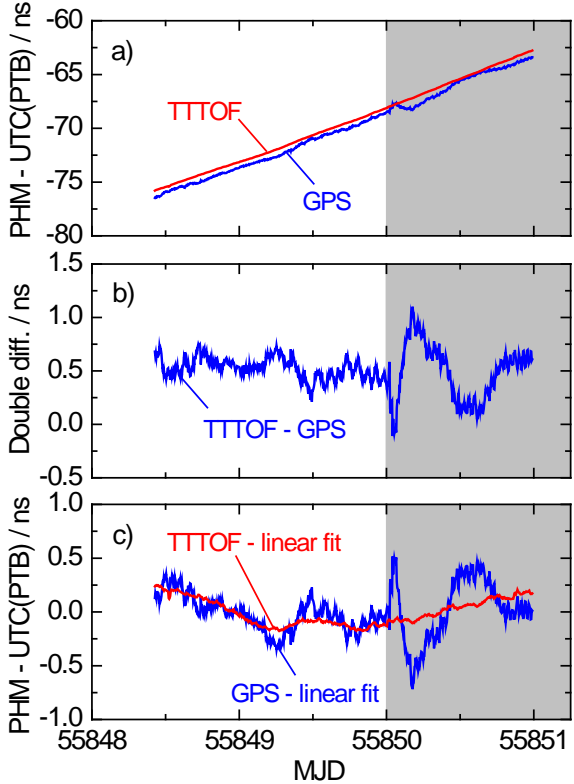

Figure 5: Results of time transfer via TTTOF and GPS, a) shows the time difference between UTC(PTB) at Braunschweig and PHM at Hannover, b) illustrates the double difference while c) shows the results with PHM frequency removed individually. MJD 55848 corresponds to 14 October 2011.

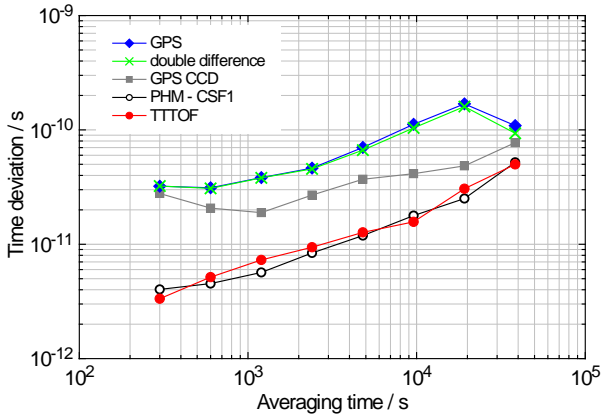

Figure 6: Time instability (TDEV): the curves represent TTTOF (red dots), GPS (blue diamonds), double difference of TTTOF and GPS (green crosses), GPS in a local common clock difference (CCD) experiment (grey squares) and local comparison of the passive hydrogen maser with PTB's fountain clock CSF1 during some days in 2008 (black circles).

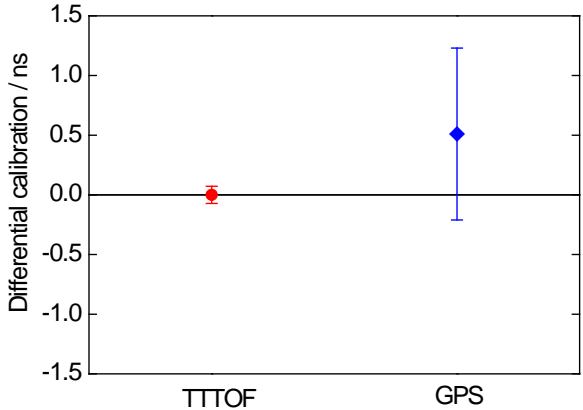

Figure 7. Illustration of the outcome of the link calibration.